\documentclass[amsmath,amssymb,aps,pre,superscriptaddress,twocolumn,showpacs,floatfix]{revtex4-1}
\usepackage[T1]{fontenc}
\usepackage{graphicx}
\usepackage{color}

\newcommand{\be}{\begin{equation}}
\newcommand{\ee}{\end{equation}}
\newcommand{\bea}{\begin{eqnarray}}
\newcommand{\eea}{\end{eqnarray}}

\newcommand{\sY}{{\sf Y}}

\newcommand{\NS}{N_\mathrm{SSS}}
\newcommand{\Nb}{N_\mathrm{b}}
\newcommand{\gammastar}{\gamma^\star}

\begin{document}


\title{Geometry and the onset of rigidity in a disordered network}


\author{Mathijs F. J. Vermeulen}
\affiliation{Department of Applied Physics, Eindhoven University of Technology, Den Dolech 2, 5600MB Eindhoven, The Netherlands}
\author{Anwesha Bose}
\affiliation{Department of Applied Physics, Eindhoven University of Technology, Den Dolech 2, 5600MB Eindhoven, The Netherlands}
\author{Cornelis Storm}
\affiliation{Department of Applied Physics, Eindhoven University of Technology, Den Dolech 2, 5600MB Eindhoven, The Netherlands}
\affiliation{Institute for Complex Molecular Systems, Eindhoven University of Technology, Den Dolech 2, 5600MB Eindhoven, The Netherlands}
\author{Wouter G. Ellenbroek}
\email{w.g.ellenbroek@tue.nl}
\affiliation{Department of Applied Physics, Eindhoven University of Technology, Den Dolech 2, 5600MB Eindhoven, The Netherlands}
\affiliation{Institute for Complex Molecular Systems, Eindhoven University of Technology, Den Dolech 2, 5600MB Eindhoven, The Netherlands}


\date{\today}

\begin{abstract}
Disordered spring networks that are undercoordinated may abruptly rigidify when sufficient strain is applied. Since the deformation in response to applied strain does not change the generic quantifiers of network architecture---the number of nodes and the number of bonds between them---this rigidity transition must have a geometric origin. Naive, degree-of-freedom based mechanical analyses such as the Maxwell-Calladine count or the pebble game algorithm overlook such geometric rigidity transitions and offer no means of predicting or characterizing them. We apply tools that were developed for the topological analysis of zero modes and states of self-stress on regular lattices to two-dimensional random spring networks, and demonstrate that the onset of rigidity, at a finite simple shear strain $\gammastar$, coincides with the appearance of a single state of self stress, accompanied by a single floppy mode. The process conserves the topologically invariant difference between the number of zero modes and the number of states of self stress, but imparts a finite shear modulus to the spring network. Beyond the critical shear, we confirm previously reported critical scaling of the modulus. In the sub-critical regime, a singular value decomposition of the network's compatibility matrix foreshadows the onset of rigidity by way of a continuously vanishing singular value corresponding to nascent state of self stress.
\end{abstract}


\maketitle

\section{Introduction}
Fibrous networks feature broadly in natural as well as synthetic materials. Examples include rubbers \cite{boyce2000constitutive}, hydrogels \cite{ahmed2015hydrogel}, and most biopolymer-based cellular and extracellular matter including the actin cytoskeleton \cite{mofrad2006cytoskeletal}, collagenous extracellular matrix \cite{fratzl2008collagen}, and fibrin-based blood clots \cite{piechocka2010structural,brown2009multiscale}. These networks are all characterized by a fairly low connectivity: In most cases, the network is held together by crosslinks that connect precisely two fibers together. Treating the crosslinking points as positional coordinates or ``nodes'', and fiber segments as bonds connecting these nodes, each node will have 4 bonds, or fewer if one discards dangling ends~\cite{broedersz2014review}.

While real fibrous networks have one or more features that makes them rigid despite this low connectivity, such as a bending stiffness or pre-stresses of mechanical, thermal and/or osmotic origin, the mechanical properties can still to a large extent be determined by the pure geometry of the network. The basic spring network that one is then left with, is governed by the arguments due to Maxwell~\cite{maxwell1864calculation} and Calladine~\cite{calladine1978buckminster}, which count degrees of freedom and constraints to determine answers to two basic mechanical questions: In in how many ways $N_0$ can the network deform without energy cost? And in how many ways $\NS$ can the network support a stress without needing external forces? In other words: How many floppy modes and states of self-stress does the network possess? Calladine's findings prove, that for every network with $N$ nodes and $\Nb$ bonds, the difference $\nu=N_0-\NS$ is the same, regardless of the spatial arrangement of its nodes and bonds. As a result, $\nu$ cannot change when the network is deformed, for instance in response to external loading. For this robustness, and the fact that $N_0$ and $\NS$ are each related to the dimensionality of certain solution spaces, $\nu$ may be regarded as a topological index, protected from spatial change~\cite{kane2013topological}. Importantly, though, the individual numbers $N_0$ and $\NS$ depend on the local details and will, generally, change during deformation.~\cite{lubensky2015phonons}. What we will demonstrate is precisely how these changes affect the transition from floppy to rigid in disordered networks.

\begin{figure}[t!]
\includegraphics[width=.23\textwidth]{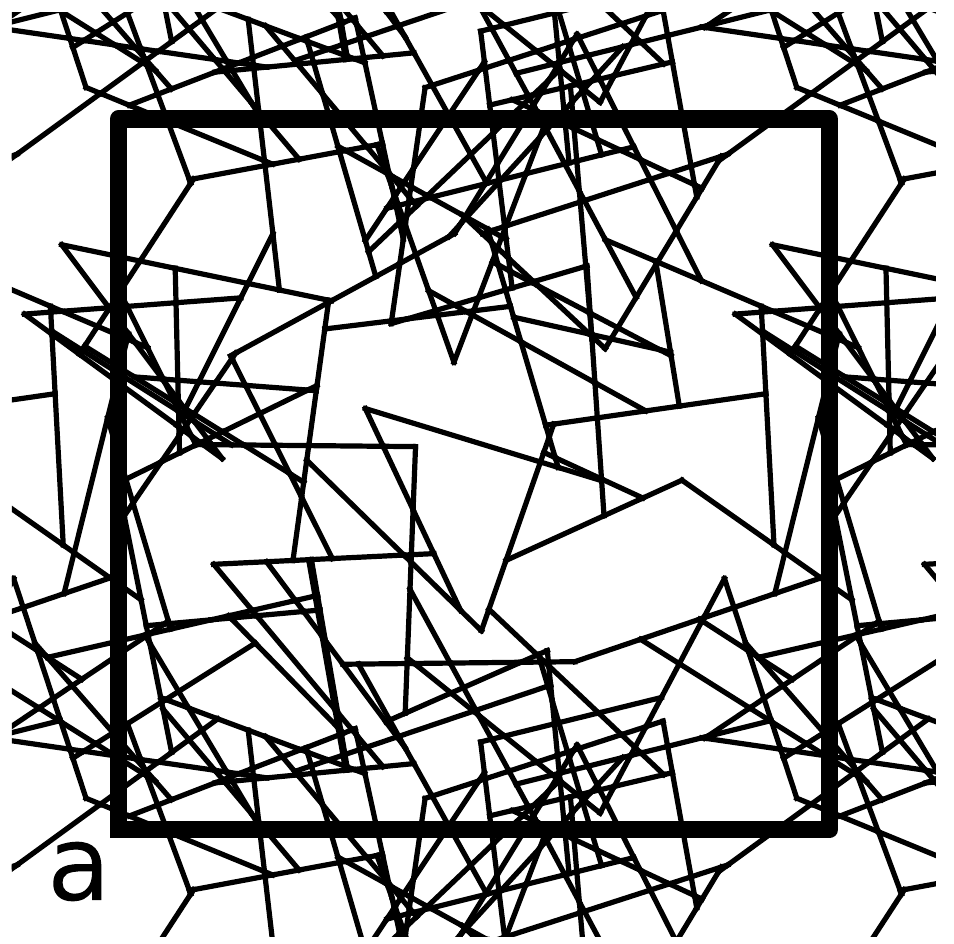}\includegraphics[width=.23\textwidth]{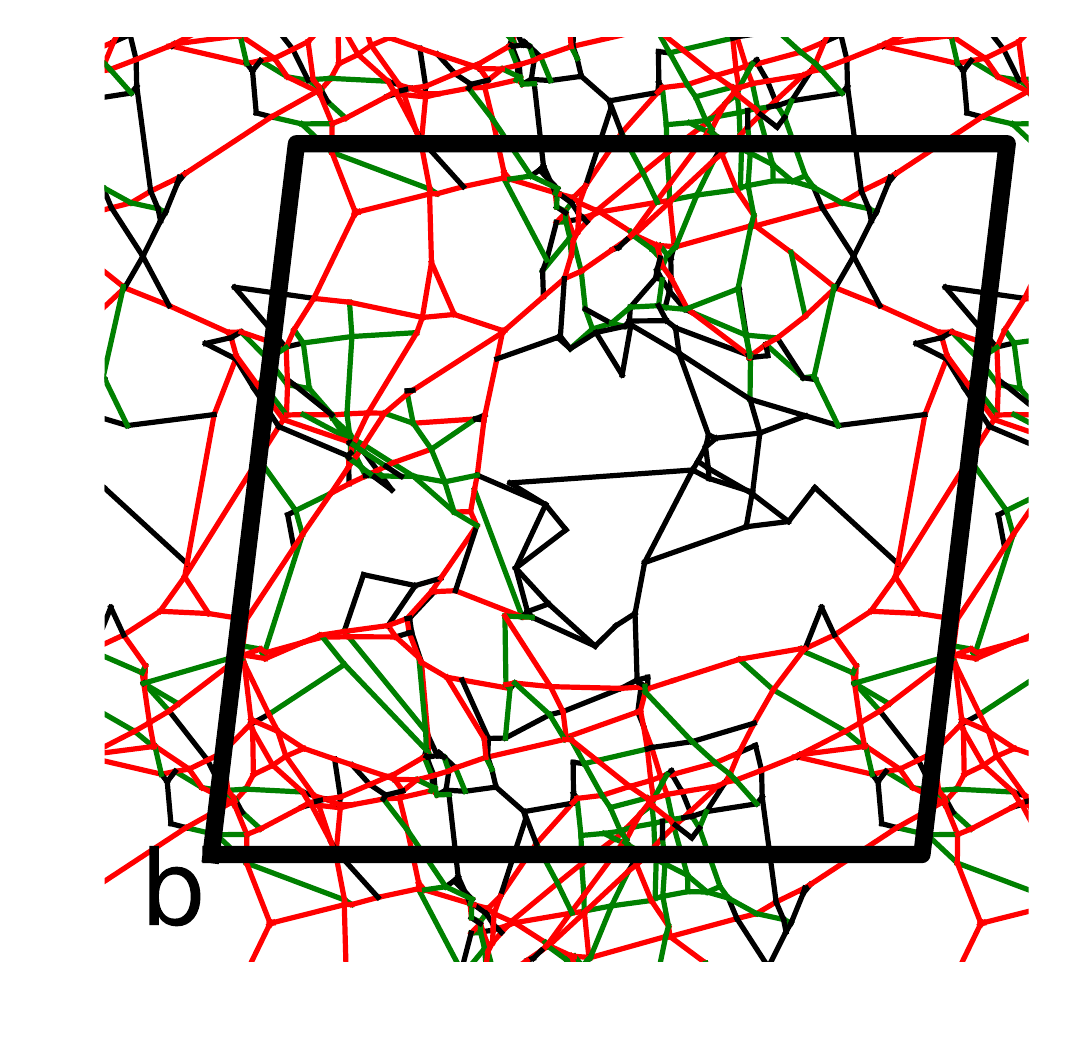}
	\caption{Floppy to rigid: (a) A network, as initially generated with the Mikado method. The periodically repeated simulation area is outlined by the thick black boundary. The average connectivity of this sample network is $z=3.6$. (b) The same network, after a simple shear strain of $\gamma=0.12$ has been applied (using Lees-Edwards boundary conditions \cite{lees1972computer}). The network is now rigid, betrayed by a single, system-spanning state of self stress comprising extended (red) and compressed (green) segments. Black segments are unstrained.}
\label{fig:network}
\end{figure}

We use the mikado model, a well-studied and often employed two-dimensional model for fibrous networks. It has an extensive number of floppy modes in the limit of zero bending stiffness~\cite{wilhelm2003elasticity}, but its states of self-stress (SSS) are a nontrivial matter that has been unexplored. The SSS are intimately related to mechanical performance: For small shear strains, these networks can deform freely, suggesting that they either do not have any SSS, or that the ones they do have are orthogonal to the shear deformation~\cite{wyartthesis,lubensky2015phonons}. At finite strain, however, these networks rigidify in a manner bearing some of the hallmarks of a continuous phase transition~\cite{wyart2008subiso,sharma2016strain,sharma2016strainPRE}. Even for networks that do have some degree of subisostatic rigidity due to fiber bending, the sharp increase at finite strain is conserved highlighting the broad relevance of the geometric rigidity transition. Rigidity can only come about when the network is capable of bearing a load, and therefore indicates the presence of at least one SSS.
The question how the SSS arises is complementary to the recent finding that the floppy modes that live on the mikado sticks can be localized to one end of the stick by slightly deforming it, similarly to how floppy modes in Kagome lattices can be localized to the boundary by perturbing the geometry~\cite{mao2017mikado}.

Our central finding is that mikado networks do not have any states of self-stress as they are created. Upon deformation, and precisely at the critical strain, a single SSS appears. This SSS is purely geometrically induced through the non-affine displacements of the nodes. The SSS are difficult to extract due to the disordered nature of mikado networks, combined with the strict geometric conditions for SSS to exist; we show that a singular value decomposition of the equilibrium matrix~\cite{pellegrino1993structural} sensitively captures the approach to, and onset of, rigidity. Finally, we validate our numerical mechanical experiment by comparing the scaling of the shear modulus beyond the critical strain with known results~\cite{sharma2016strain,sharma2016strainPRE}.

\section{The mikado model: Definition and mode structure} \label{sec:model}

The mikado model is often used to study the mechanical properties of disordered semi-flexible fibrous or polymer networks \cite{vrusch2015curvature,wilhelm2003elasticity,cioroianu2013normal,broedersz2012filament}. Straight, segmented fibers (``sticks'') are placed randomly in a doubly periodic two-dimensional simulation box. In our work, the simulation box is a square with side $L$ during the mikado procedure, but we will deform the networks later by changing the shape of the periodic box using a simple shear strain $\gamma$, as illustrated in 
 Fig.~\ref{fig:network}. The coordinates of the intersection points of the fibers (the ``crosslinks'', or ``nodes'') are the degrees of freedom of the model. Interactions between nodes that are connected by a stick capture the mechanical response of the sticks. The original mikado model, for instance, energetically penalizes both the stretching and the bending of a stick; in this paper we shall focus on the limit of vanishing bending energy. The network is thus a central force (spring) network. The energy of the model is given by
	\begin{equation}\label{eqn:Hamiltonian}
	\mathcal{E}=\frac{1}{2}\sum_{i=0}^{N_{\mathrm{b}}} \frac{\sY}{\ell_{0,i}}\left(\ell_i-\ell_{0,i}\right)^2~.
\end{equation}	 
Here, $\sY$ is the Young's modulus of a single fiber, and $\ell_{0,i}$ is the rest length of the $i^{\rm th}$ spring. The spring constant of spring $i$ is therefore $k_i=\sY/\ell_{0,i}$. $\ell_i$ is the instantaneous length of spring $i$ spring, which is a  function of the coordinates of the crosslinks. Specifically, for a network with $N$ nodes and $\Nb$ springs, we may collect the deviations of all crosslinks from their initial (unstrained) positions into the $2N$-dimensional displacement vector $\mathbf{U}$, which is mapped onto the $\Nb$-dimensional vector $\mathbf{E}$ containing all spring elongations by the $\Nb \times 2N$ compatibility matrix $\mathcal{C}$:

\begin{equation}
	\mathcal{C}\mathbf{U}=\mathbf{E}.
\end{equation}

A floppy or zero mode, now, is a set of node displacements that costs zero energy, {\em i.e.} that does not change the length of any of the bonds. That is, the vector $\mathbf{E}$ contains only zeroes. Floppy modes may this be determined as vectors in the null space of the compatibility matrix $\mathcal{C}$. As a result, the number of floppy modes, $N_0$, equals the nullity (the dimension of the null space) of $\mathcal{C}$. 

A state of self stress is a configuration of tensions in all of the springs satisfying mechanical equilibrium. That is, the sum of forces $\sum {\mathbf f}={\mathbf 0}$, on each node separately. Mapping from the tensions in the springs to the forces on the nodes is effected by the $2N \times \Nb$ equilibrium matrix $\mathcal{Q}$. It is a function of the positions of the nodes, and hence, indirectly of the shear strain $\gamma$ on the network. Collecting all bond tensions in a $\Nb$-dimensional vector $\mathbf{T}$, and all spring forces ${\mathbf f}$ into the $2N$-dimensional vector $\mathbf{F}$, the condition for mechanical equilibrium may be expressed as
	\begin{equation} \label{eqn:qt0}
	\mathcal{Q}\mathbf{T}=-\mathbf{F}=\mathbf{0}.
	\end{equation}
Non-trivial solutions $\mathbf{T} \neq \mathbf{0}$ to this equation are called states of self stress; the number of states of self stress $\NS$ is equal to the nullity of $\mathcal{Q}$~\cite{sun2012surface}.


Our main numerical results concern networks that have been deformed by changing the shape of the periodic simulation box. In order for this to represent a quasistatic deformation, we minimze the energy, Eq.~(\ref{eqn:Hamiltonian}), after each box shape change, using the conjugent gradient algorithm \cite{press1989numerical}. In networks that have been deformed beyond the critical strain, we determine the components of the virial stress
\be
\sigma_{ij}=\frac{1}{2 L^2}\sum_{\langle kl \rangle}({\mathbf x}_{k,i}-{\mathbf x}_{l,i}) {\mathbf f}_{kl,j}\, ,
\ee  
with $i,j=x,y,x$, $L$ the box size, $\langle kl \rangle$ shorthand for summation over all node indices $k,l$ connected by a spring, ${\mathbf x}_{k,i}$ the $i$-component of the coordinates of node $k$ and ${\mathbf f}_{kl,j}$ the $j$-component of the force along the spring connecting nodes $k$ and $l$. The shear modulus is then computed as
\be
G(\gamma)=\frac{\partial \sigma_{xy}}{\partial \gamma}\,,
\ee
which we evaluate in linear response using the full Hessian (as the network will be stressed when $\gamma>\gammastar$)~\cite{ellenbroek2009jammed}.

Maxwell-Calladine counting dictates that the difference between the number of floppy modes and the number of states of self-stress is fixed by the count of degrees of freedom and constraints,
	\begin{equation} \label{eqn:maxcall}
		\nu= N_0-\NS=2N-\Nb-3~,
	\end{equation} 
where the right-hand side can be understood to represent the number of degrees of freedom (the $2N$ node positions), minus the number of constraints imposed by the springs (1 for each of the $\Nb$ springs), minus the number of trivial zero modes (in two dimensions, 2 translations and 1 rotation)~\cite{calladine1978buckminster,lubensky2015phonons}. 
Thus, the value of $\nu$ can be obtained simply by counting nodes and springs, but to find $N_0$ and $\NS$ separately requires more effort. As long as the geometry of the network is completely generic, the pebble game algorithm can be used to keep track of the redundant bonds~\cite{jacobs1995generic,jacobs1996generic}. This algorithm is sensitive to local properties, while the Maxwell-Calladine argument only gives global information. The pebble game is, however, still topological in the sense that it takes as input only which node is connected to which other nodes, but is otherwise blind to geometric happenstance. Mikado networks transition from soft to rigid upon deformation, and this transition does not change their topology. This belies a geometric origin: beyond the critical strain, there is a state of self-stress that was not there earlier, even though no bonds were created or destroyed during the deformation. Both the Maxwell-Calladine count and the pebble game give the same result, below and beyond the critical strain. 
The only way to reveal the actual floppy modes and states of self-stress of these networks is, therefore, to use an algebraic analysis that is sensitive to the geometric details underpinning the rigidity transition.  

Eq.~(\ref{eqn:maxcall}) suggests that mikado networks have an extensive excess of zero modes relative to their states of self stress; $\nu \sim N_\mathrm{sticks}$. To see this, note that in a mikado network, most nodes have 4 bonds connecting to them because they are the intersection of two lines, so that the number of bonds should be $\Nb=2N$ (dividing by 2 to avoid double-counting). However, each mikado stick has two dangling ends which do not contribute to the network, so $\Nb=2N-N_\mathrm{sticks}$ (each dangling end corresponds to half a bond). Thus $N_0-\NS=N_\mathrm{sticks}-3$. Alternatively, \cite{mao2017mikado} points out that the addition of each stick creates a floppy mode corresponding to the longitudinal displacement of the new stick. In dilute networks, the actual tally of floppy modes can be a bit lower than the number of sticks.

\begin{figure}[hb]
\includegraphics[width=.5\textwidth]{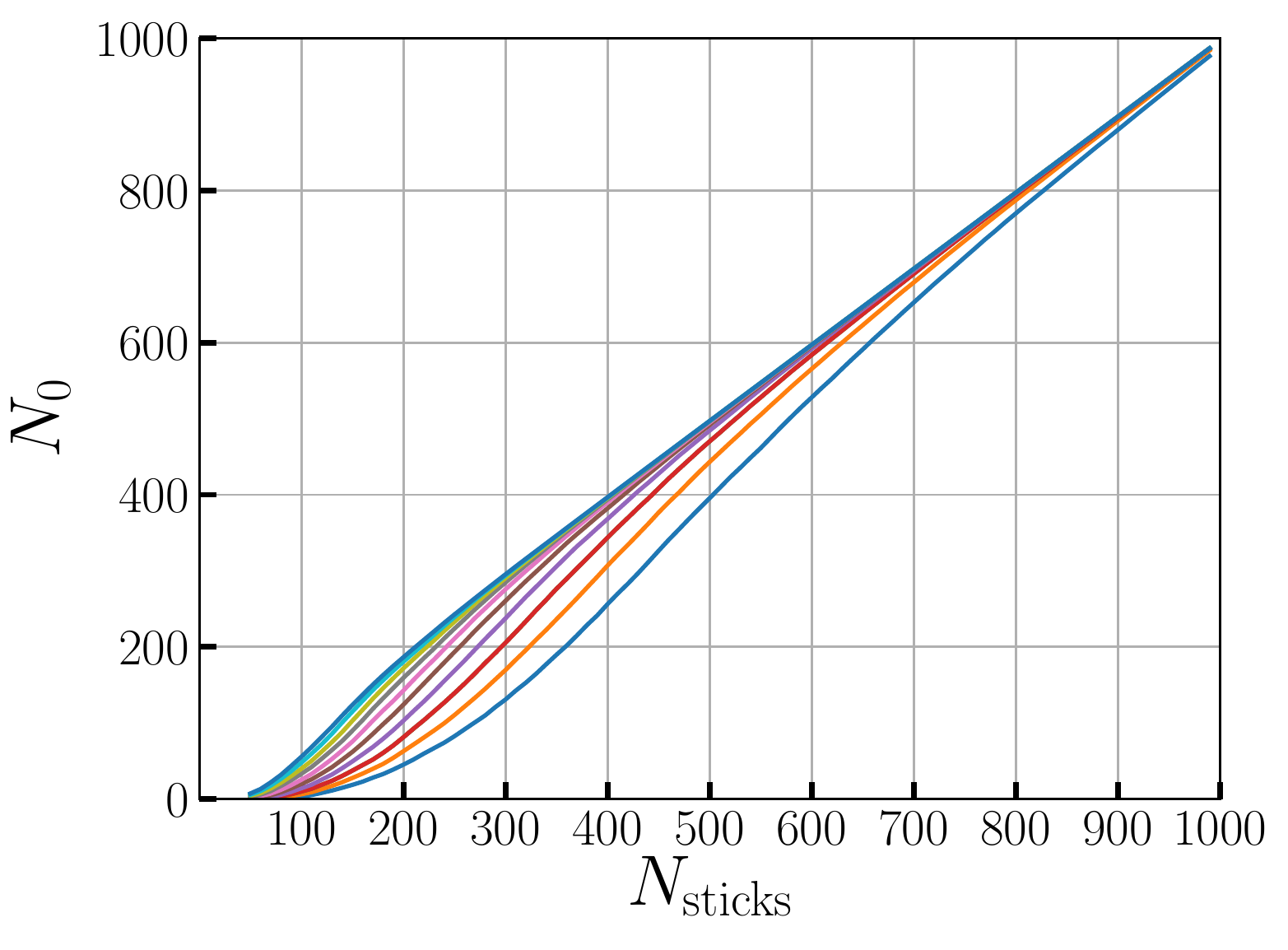}
	\caption{The number of floppy modes is counted as a function of the number of mikado sticks (fibers) used, for a range of stick lenghts. In the limit of long fibers, or a high density, the number of floppy modes scales with the number of fibers in the network. The curves are ordered bottom-to-top from short fibers (10\% of the box size) to longer fibers (19\% of the box size).}
\label{fig:nofloppymodes}
\end{figure}

As an illustration, we calculate the number of floppy mode for mikado networks as a function of the number of fibers, for different fiber lengths. The results are shown in Fig.~\ref{fig:nofloppymodes}. Initially, all the curves deviate from the predicted linear asymptote. This is a low-density effect, the fibers are deposited at a random position and orientation and a newly added fiber may not intersect any other fiber, adding no degrees of freedom in the model. At higher densities, all lines converge to the predicted asymptote to the same line, showing that indeed each added fiber contributes one zero mode: $N_0=N_\mathrm{sticks}$.

While the excess is thus decidely extensive (and large mikado networks are certainly riddled with floppy modes), we cannot yet say anything about the density-dependence of $N_0$ and $\NS$ separately; we now investigate the partitioning of $\nu$.

\section{Connectivity and Rigidity}
\begin{figure}[hb!]
\includegraphics[width=.3\textwidth]{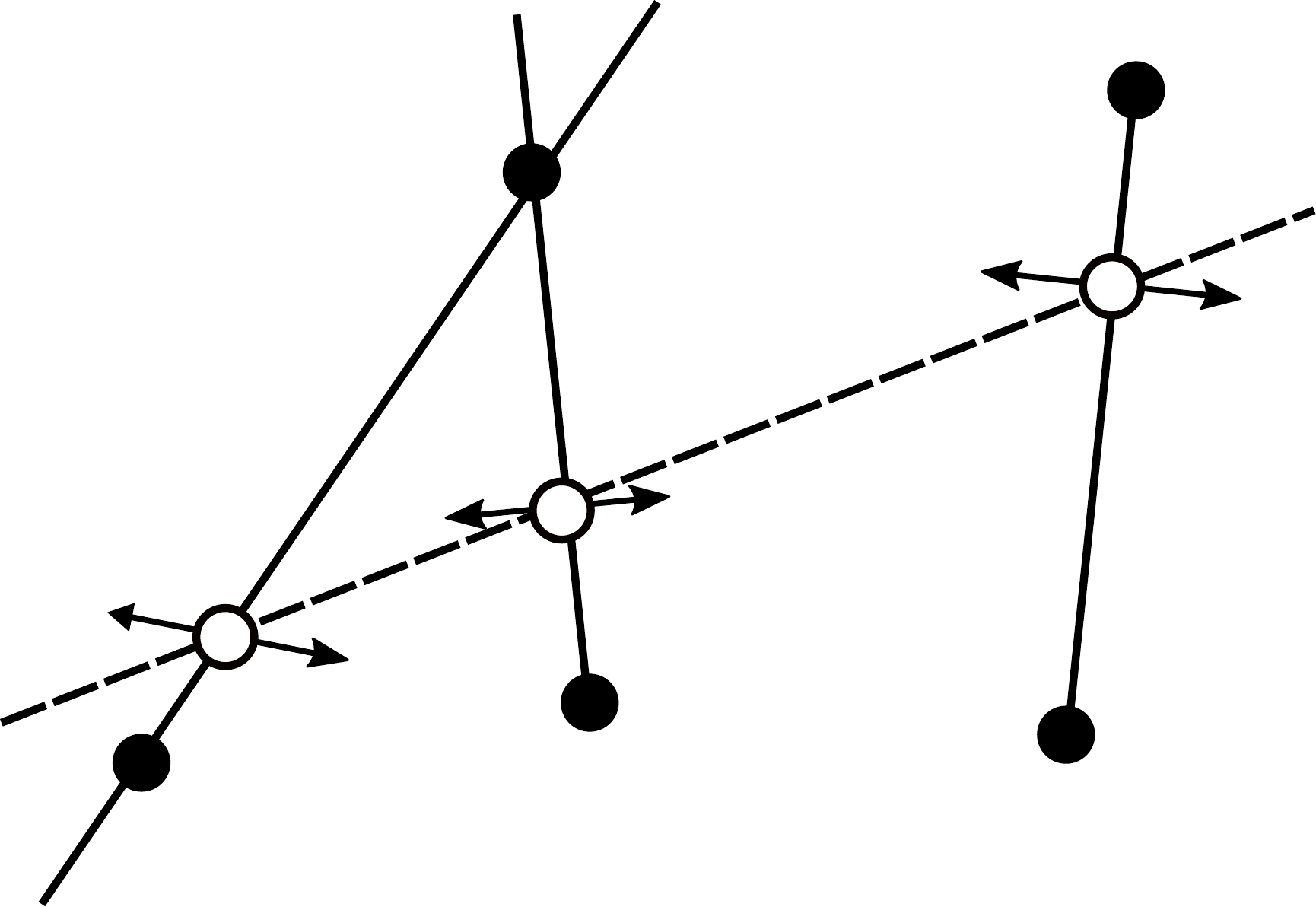}
	\caption{The addition of a mikado stick to an already existing network gives rise to $\mu$ (in this example $\mu=3$) intersections and $2\mu-1$ new springs. Turning the intersection points into nodes before adding the springs that represent the new stick creates a floppy mode at each new node, indicated with the arrows. Adding the $\mu-1$ new springs couples these into a single floppy mode.}
\label{fig:mikadnew}
\end{figure}

Getting straight to the point: Unstrained mikado networks do not have any states of self-stress. To see this, consider what happens when a single mikado stick is added to an existing network, as sketched in Fig.~\ref{fig:mikadnew}. This increases the topological index $\nu$ defined in Eq.(~\ref{eqn:maxcall}) by one---the newly added fiber adds $\mu$ (the number of previously present fibers it crosses; in this example, 3) new nodes to the network, and adds $2\mu-1$ new springs. These new springs are $\mu-1$ springs that connect the new nodes, and another $\mu$ new springs connecting previously existing nodes. The right hand side of equation \ref{eqn:maxcall} has therefore increased by one. Thus, at least one zero mode has been created---possibly more, but then these should be accompanied by states of self stress; $\Delta N_0=\Delta \NS+1$. One new floppy mode is readily identified, and consists of an infinitesimal, coherent motion of the newly created nodes as indicated in Fig.~\ref{fig:mikadnew}. 

It turns out, that $\Delta \NS=0$, and thus that exactly one floppy mode is created by adding a new stick. To see this, consider the force balance equations on any of the newly created nodes. Because the two fibers that intersect at this node are both straight, these equations can be decomposed into components along the fibers. Firstly, the forces along the new fiber must all be equal, and because the fiber has finite length (there is a node with only 3 springs connecting to it at each end) they must all be zero. More importantly, the forces in the two springs in which the existing spring was divided are equal, so that nothing has changed about the possible force networks on the existing network. The conclusion is that adding a mikado fiber to any network does not create any states of self-stress and adds precisely one floppy mode. By induction, a network that consists solely of mikado fibers does not have any states of self-stress and will, in general, not support a mechanical load.

\begin{figure}[hb]
\includegraphics[width=.5\textwidth]{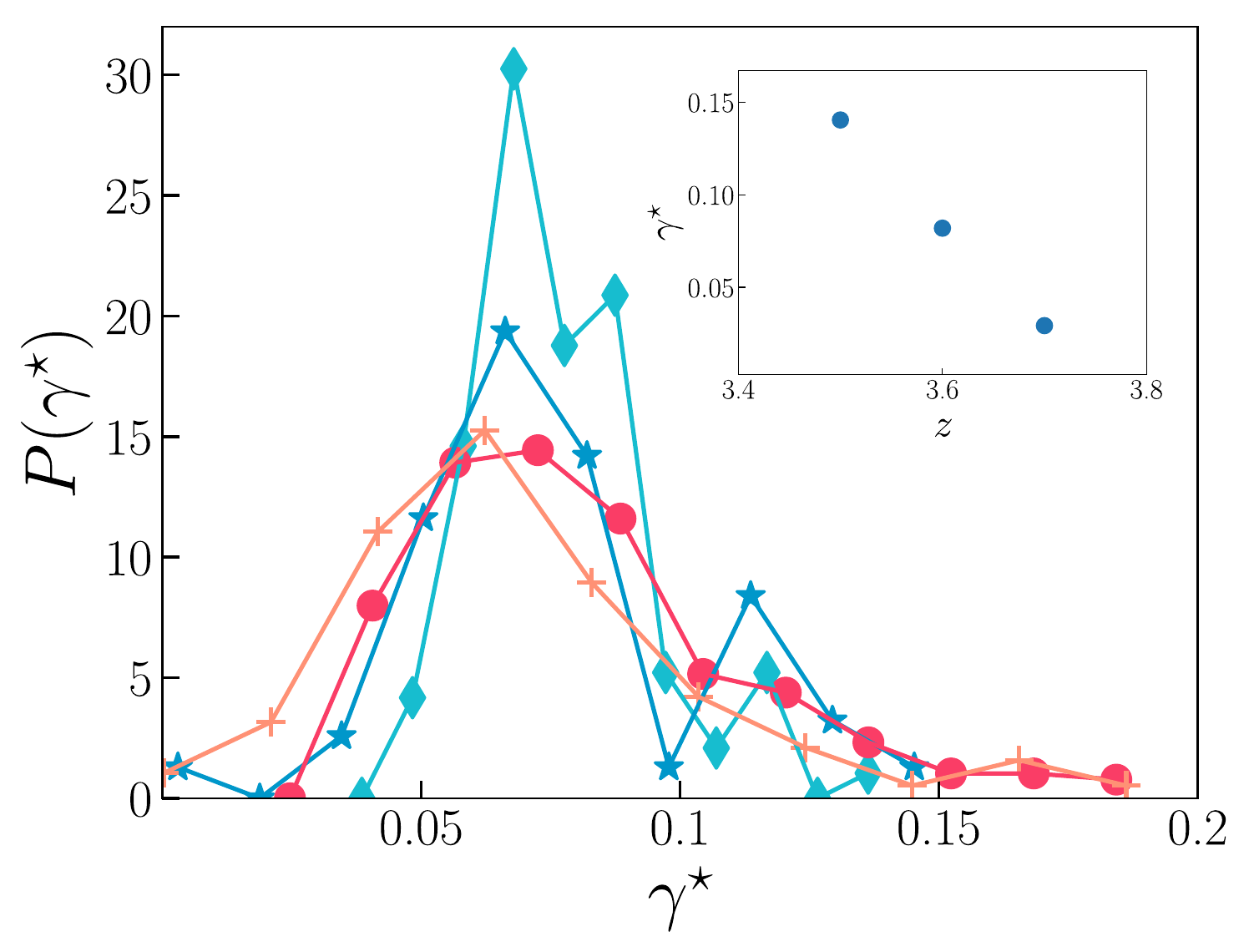}
	\caption{Probability density of the critical strain $\gammastar$, the strain at which the network acquires a modulus, for mikado networks at $z=3.6$ with stick length $0.57$. System sizes are $L=1 (+) ,\sqrt{2} (\bullet) ,2 (\star) ,2\sqrt{2} (\blacklozenge)$, for decreasing distribution width.   The inset shows the average $\gammastar$ as a function of connectivity $z$ (calculated at $L=\sqrt{2}$).}
\label{fig:sketchmod}
\end{figure}
Large deformations, however, may rigidify a
mikado network: without changing the network topology, a state of self-stress appears. We will now explore how this happens,
focusing on simple shear (other modes of deformation such as global stretching will produce similar phenomena). 
As a preliminary, we note that the critical shear strain $\gammastar$ at which rigidity sets in depends primarily on the coordination number of the network, becoming more narrowly distributed as the system size increases. Fig.~\ref{fig:sketchmod} shows this distribution, for a range of system sizes, keeping the connectivity fixed at $z=3.6$. Thus, the notion of a $z$-dependent critical strain becomes tightly defined in the large-system size limit of the mikado model. Beyond this  
critical strain $\gammastar$, since the boundaries are periodic (Lees-Edwards), there can be no external forces on any of the nodes. Thus, the force network comprising the stretched and compressed springs giving rise to the finite modulus represents a state of self-stress. We emphasize, again, that this state of self-stress cannot be found by purely topological methods such as the pebble game.

Instead, we must directly compute the non-trivial solutions of Eq.~(\ref{eqn:qt0}). A reliable and numerically stable way to establish these is to use a singular value decomposition \cite{riley2006mathematical},\cite{pellegrino1993structural}, \cite{pellegrino1986matrix}, decomposing the equilibrium matrix $\mathcal{Q}$ as

\begin{equation}
	\mathcal{Q}=\mathcal{A}\, \mathcal{S} \, \mathcal{B}^{\dagger},
\end{equation} 
where the $\mathcal{A}$ ($2N\times 2N$) and $\mathcal{B}$ ($\Nb\times\Nb$) are orthogonal matrices, the dagger denotes a Hermitian conjugate, and $\mathcal{S}$ is a $2N\times \Nb$ rectangular diagonal matrix, holding as its $\Nb$ main-diagonal entries $(\mathcal{S})_{ii}$ the so-called singular values $s_i$. If any of those singular value is equal to zero, the corresponding column of $\mathcal{B}^{\dagger}$ is in the kernel of the equilibrium matrix $\mathcal{Q}$ and thus a state of self-stress of the system. 

\begin{figure}[h]
\includegraphics[width=.4\textwidth]{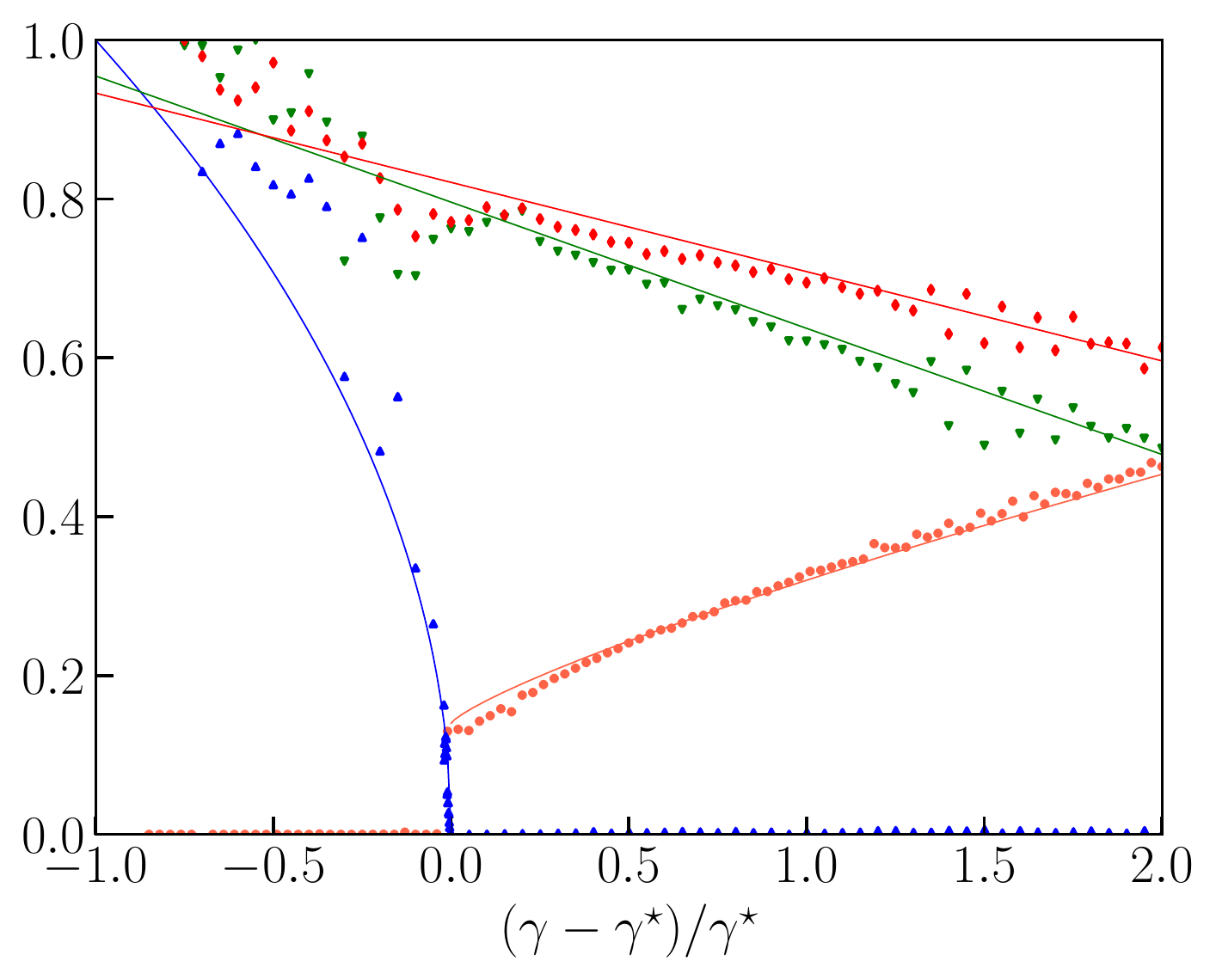}
	\caption{The scenario for shear-induced rigidification of the mikado network: as the critical strain $\gammastar$ is approached, one singular value drops to zero spawning a state of self stress that carries the load, prompting a finite shear modulus. We plot the lowest singular value in blue, the modulus in red, and the second and third lowest singular values in green and red, respectively. It is evident, that even well beyond the transition there remains only one state of self stress. The line through the shear modulus graph is a fit to Eq. (\ref{eqg}), the other lines are guides to the eye.}
\label{fig:rheology}
\end{figure}

We performed a series of numerical simple shear experiments on an ensemble of mikado networks with a an average connectivity of $z=3.6$, and fibers of initial length $\frac25\sqrt2\approx 0.57$ in box sizes ranging from $L=1$ to $L=2\sqrt2$. The number of mikado sticks involved ranges from order 50 for the smallest system to order 400 for the largest one.
We numerically minimize the network energy, Eq.(\ref{eqn:Hamiltonian}) after each shear step and compute the shear modulus, the fraction of springs that carries a nonzero tension, and the three lowest of singular values of the equilibrium matrix. The result of these calculations is shown in Fig.~\ref{fig:rheology}, as a function of the strain.

\begin{figure}[h]
\includegraphics[width=.4\textwidth]{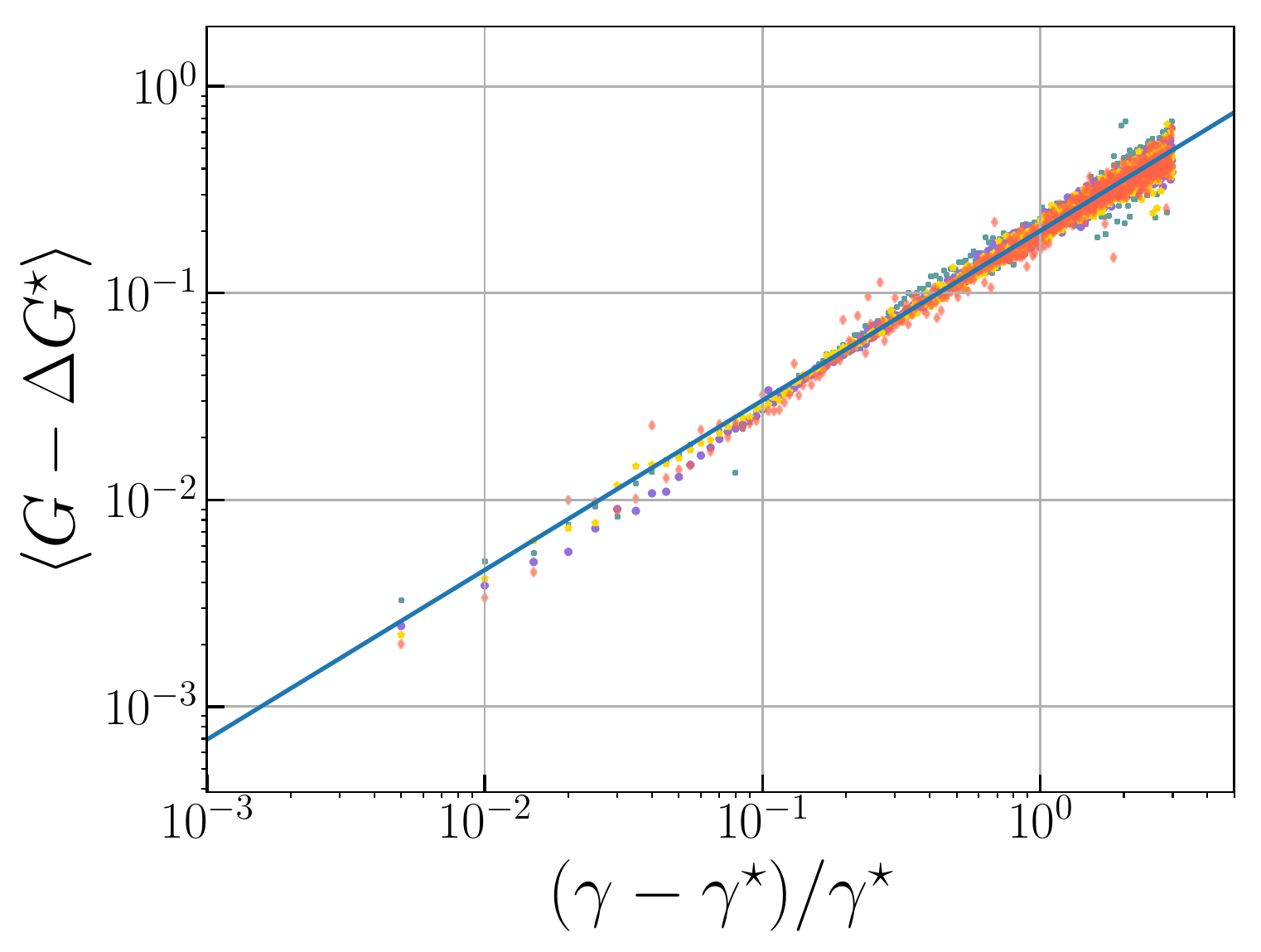}
	\caption{Beyond the critical strain, the shear modulus rises as a power law; $G-\Delta G^\star \sim \left[(\gamma-\gammastar)/\gammastar\right]^f$ with $f=0.82$.}
\label{fig:powerlaw} 
\end{figure}

This shear modulus jumps at the critical strain, in apparent contradiction to earlier reports that the transition is continuous \cite{sharma2016strain}. To analyze this effect, we fit the measured moduli to the following equation
	\begin{equation}\label{eqg}
	G(\gamma)=\Delta G^\star+\left(\frac{\gamma-\gammastar}{\gammastar}\right)^f\, 
	\end{equation}
allowing to simultaneously quantify the post-rigidification behavior and the jump. Fig. \ref{fig:powerlaw} and the inset summarize the scaling behavior: beyond the critical strain, the modulus rises according to a power law, with an exponent $f=0.82$, in agreement with earlier findings for the 2D mikado model without bending. From the same fits, we have determined that the jump $\Delta G^\star$ decreases as the system size is increased, but whether it disappears completely at $L\to\infty$ is unclear; whether the jump is a real effect or an artifact of our method of determining $G$ very close to the transition requires systems too large for our current simulations to handle.

The overall scenario for the emergence of rigidity at finite strain is collected in Fig. \ref{fig:rheology}. In the deeply subcritical regime, there are no states of self stress and the system is floppy with respect to shear. Approaching the critical strain, one of the singular values drops steeply. When it touches zero, a significant fraction of the network is engaged mechanically and the shear modulus begins to rise in power law fashion. As far as we have been able to discern, no secondary states of self stress are created, and the mechanical response is thus completely carried by the first SSS to appear.

\section{Discussion}
We have shown, that the lowest singular value of the equilibrium matrix vanishes at the critical strain. At this same strain a state of self stress emerges. Of course, this event also coincides with the emergence of a finite modulus. Across the ensemble of random networks we generated, we have only ever seen a {\em single} state of self stress being responsible for the onset of rigidity. Moreover, this mode continues to be the sole mode supporting the load for some range of strains; the second and third smallest singular values remain nonzero well beyond the transition.

This finding suggests that in order to understand the striking similarities between induced rigidity and critical transitions in Ising-like models in the presence of aligning fields (evidenced, for instance, by the Widom collapse of the scaled modulus before and after the critical strain in \cite{sharma2016strain}) we should direct our attention to the properties of the first critical state of self stress, as it is the only structure responsible for the finite modulus. Its contribution to this modulus may be split up into two parts, one having to do with the springs (providing a constant contribution proportional to the spring constant per spring) and a contribution proportional to the spring tension \cite{alexander1998review,ellenbroek2009jammed}; a closer look at the statistical properties of the critical SSS (participation ratio, tension distribution) will likely offer a deeper understanding of the scaling near the rigidification threshold. 

Geometric rigidification offers great advantages over architected rigidity. These are materials that are initially vanishingly soft and malleable; the application of a load rigidifies them. There is likely an intimate connection between the critical SSS and the strain that prompted it; in shear, the SSS brings a finite shear modulus. In uniform dilation, it will produce a nonzero bulk modulus. In other words, the material becomes resilient to the strain that prompted rigidity (but not necessarily against other modes of deformation). Because tailored resilience may be coaxed out of very sparse network geometries, protocols leveraging this effect may be relevant to produce generic, lightweight architectures for predefined mechanical performance.
\section*{Acknowledgments}
We are grateful to Stefan Paquay for technical advice. This work is part of the research programme on Marginal Soft Matter (FOM12CSM01), which is financed by the Netherlands Organisation for Scientific Research (NWO).

\end{document}